# Atomic collider into dual-isotope magneto-optical trap

Gregory Surdutovich

Institute of Laser Physics Siberian Branch of Russian Academy of Sciences
13/3 Lavrentyeva Pr., 630090, Novosibirsk, Russia.

**ABSRACT**

When two of three pairs of the Gaussian laser beams of a traditional MOT are misaligned in the "racetrack" configuration the effective coordinate-dependent vortex force do arise. Then an atom is accelerated by this vortex force until its velocity not balanced by the damping force. This situation may produce a stable ring of revolving atoms of a certain radius. Due to the different frequency and laser beams intensity dependences of the vortex, damping and trapping forces it is possible to equalize the radii of two orbiting groups of atoms in two-species or dual-isotope magneto-optical trap and so to arrange a continuing collider of cooled atoms with the prescribed relative velocity. A collider setup for atoms of two different types rotating with different angular velocities along the same ring-like trajectory into MOT of the conventional six-beam geometry is proposed and designed on example of two rubidium isotopes $Rb^{85}$ and $Rb^{87}$.

## 1. Introduction

Laser cooling and trapping of atoms in MOT is a powerful method to produce a large number of atoms at very low temperature and investigate their interactions. During development of such techniques, it became evident that atomic collisions were one of the main limiting factors in achievement of high density samples on the road to Bose-Einstein condensation since the loss mechanisms were caused by laser-induced collisions.

On the other hand, the laser-cooling technique could be used as a tool to study collision processes at energy range by more than six orders of magnitude lesser in comparison with the thermal collisions. A large fraction of atoms in MOT are in the excited state and so the excited-state collisions play an important role. It should be mentioned as well that their cross section may be many orders of magnitude larger than for ground-state collisions.

Long duration excited- state collisions between slow particles in presence of a light field became a fruitful subject from which high-resolution information on molecular structure could be obtained. For typical temperatures in optical traps, the velocities of the atoms are sufficiently low so that a spontaneous decay may take place before or in the process of collision. Therefore, the collision dynamics is very sensitive to tuning of the laser frequency. Last years achievements in construction of the duel-isotopes [1-3] (DIMOT) and two-species [4-6] (TSMOT) magneto-optical traps open new possibilities for cold and ultra-cold collisions studies and photo-associative spectroscopy. Here I propose use specific spatial ( rotating ring-like) modes of different atoms or isotopes for design of a circular atomic collider for cold atoms with the calculated velocity and a given level of excitation.

## 2. Calculation of the orbits in conventional misaligned MOT

Ring-like spatial distributions (modes) of atoms orbiting around a core were firstly observed in a misaligned cesium MOT [7] and explained in terms of the conventional MOT forces acting on each individual atom plus the assumption about influence of the collective interatomic forces acting between the trapped atoms [7,8]. After observation in sodium MOT the variety of spatial structures of cooled atoms (including coreless rings),[9], a simple model of coordinate-dependent vortex forces was developed which allowed to explain all observed cooled atoms structures and the transitions between them in terms of forces acting on each individual atom [10,11].

Email: Gregory@laser.nsc.ru

Let us consider the misaligned in the xy plane trapping laser beam configuration as shown in Fig. 1(a) with the perfectly aligned beams in the z direction. We assume all beams as ideal Gaussian functions, so that the Rabi frequency $V$ for each one can be written as $V = V_0 \exp(-r^2/2w^2)$, where $w$ is much larger than the wavelength.

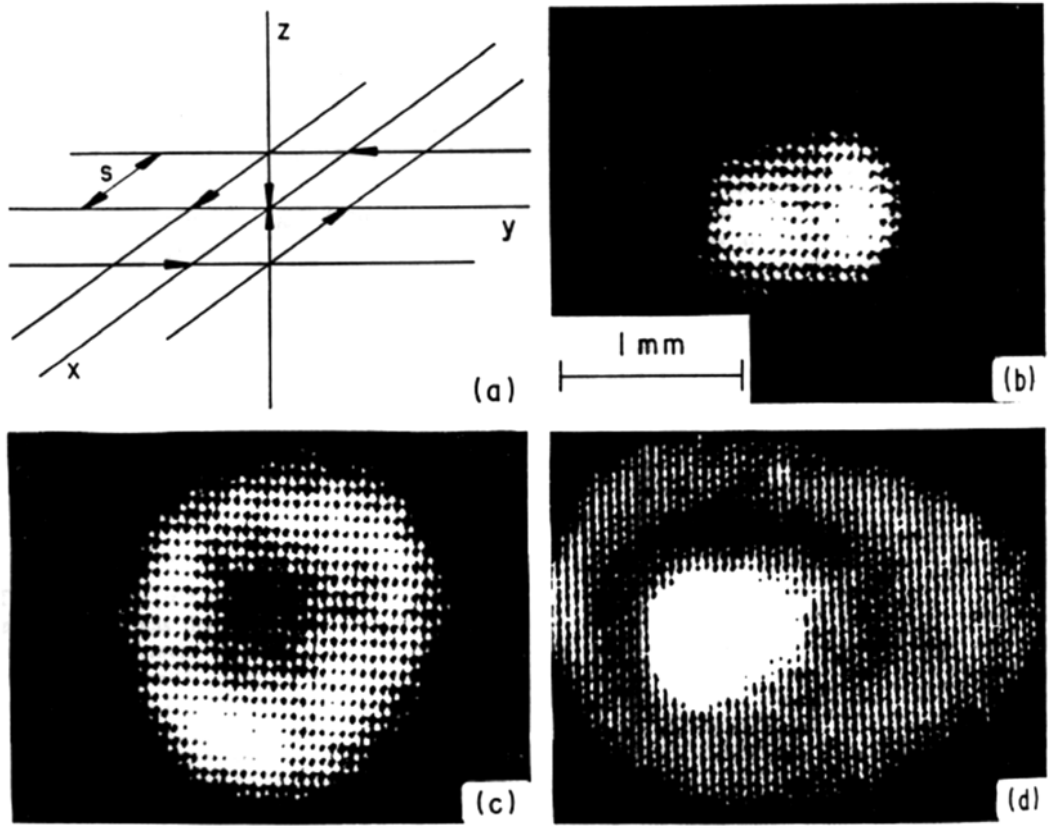

Fig.1. (a) Geometry of the laser beams forming a racetrack in the xy plane and exactly aligned in the z direction. The shift of each beam from center of a trap is labeled as $s$.
(b) Spatial atomic distribution for small misalignment and/or small detuning or high magnetic field gradient. (c) Moderated misalignment, sufficiently large detuning and the magnetic field gradient. (d) Same as in (c), but with higher beams intensity (saturation) and smaller detuning.
(adopted from Ref. [10]).

Due to the misalignment, the radiative force acting on the atom along the y direction has an x dependence and vice versa. In other words, beside the velocity and field-intensity dependent terms in the force expression an extra azimuthal component do appears, which is referred to as the vortex force. It is clear from consideration of the forces in xy plane.
For a Gaussian beam propagating exactly along the x-direction, the velocity-independent part of the radiative force has the form [12,9]

$$\vec{F} = \frac{4\hbar k \Gamma}{(1 + 4\Delta^2 + 2V_0^2)^2} V_0^2 \exp(-\frac{y^2}{w^2})\,\vec{x} \qquad (1).$$

Here $k = 2\pi/\lambda$ is the magnitude of the wave vector, $\Gamma$ is the natural width of the transition. The detuning $\Delta$ and the Rabi frequency $V_0$ are given in units of the linewidth $\Gamma$. For the beam propagating along the y-direction we have analogous expression with juxtaposition $x \leftrightarrow y$. These two orthogonally oriented coordinate-dependent harmonic forces correspond to two coordinate-dependent parabolic potential wells along x and y directions. It means that the effective potential well and the effective restoring (trapping) force have azimuthal dependence with the period $\pi/2$. This restoring force is proportional to the gradient $\nabla \vec{B}$ of the magnetic field $\vec{B}$ regulating the light pressure. The Doppler shift in the restoring force provides a velocity-dependent damping force that makes the atomic motion in MOT to be strongly overdamping [8,9]. The azimuthal dependence of the damping (friction) force is identical to the dependence of the trapping force.

Misalignment of the laser beams (Fig. 1(a)) makes the trapping potential to be slightly nonharmonic and (what is much more important) introduces the vortex force.

### The model of an isotropic harmonic trap

Neglecting by all angular dependences (in the limit $x, y \ll w$ and $s \ll w$) one comes to the most simple model of an isotropic harmonic trap with the radial-dependent vortex force [9, 10]. The radial dependence of the vortex force includes the linearly increasing part and nonlinearly decreasing term which connected with value of the laser beams waist $w$ [9]. It is convenient to describe the atomic motion in xy plane by use of the complex variable $Z = x + iy$. As a result, for an atom of the mass $m$ equation of the motion takes the form

$$m\ddot{Z} = -\alpha Z - \beta \dot{Z} - i\xi(r) Z \qquad (2).$$

Here first, second and third terms in the right-hand part of this equation correspond to the trapping, friction and vortex forces, respectively. Trapping constant is

$$\alpha = \frac{16\hbar k \Gamma |\Delta|}{(1 + 4\Delta^2 + 2V_0^2)^2} V_0^2 \frac{d\omega_0}{dx} = \tilde{\alpha} \frac{d\omega_0}{dx} \qquad (3),$$

where $\omega_0$ is the frequency of the working transition modified by the magnetic field. In this approximation friction coefficient $\beta$ is related to $\tilde{\alpha}$ as $\beta = \tilde{\alpha} k$.

$$\xi(r) = \frac{s\Gamma}{|\Delta| w^2} \tilde{\alpha}(1 - \frac{r^2}{w^2}) = \xi_0 (1 - \frac{r^2}{w^2}) \qquad (4)$$

Seeking the solution of Eq. (2) in the form $Z = Z_0 \exp(i\Omega t)$ we get

$$\Omega = \Omega' + i\Omega'' = \Omega_\pm = -\frac{\beta}{2m} \pm \sqrt{\left(\frac{\beta}{2m}\right)^2 - \frac{\alpha + i\xi(r)}{m}} \qquad (5)$$

To obtain stable ring solution (closed trajectory) one should guarantee equality $\Omega' = 0$, i.e.

$$\beta \sqrt{\frac{\alpha}{m}} = \xi(r) \qquad (6)$$

Using Eq.(4) for the vortex force we get expression for stable ring radius in this model [9]

$$r = w\left(1 - \sqrt{\frac{\alpha}{m}\frac{\beta}{\xi_0}}\right)^{1/2} = w(1 - B\varphi(I,\Delta))^{1/2} \qquad (7).$$

Here $B$ is a numerical const which depends only on the misalignment, gradient of the magnetic field and mass of an atom. The function $\varphi(I,\Delta)$ describes of the ring radius dependence in terms of unitless field intensity $I = V_0^2$ and detuning $\Delta \equiv |\Delta|$:

$$\varphi(I,\Delta) = \frac{I\Delta^{3/2}}{(1 + s_0 I + 4\Delta^2)^2} \qquad (8)$$

Here the parameter $s_0$ in the denominator describes the chosen model of saturation ($s_0 = 2$ if take into account saturation from only one pair of laser beams and $s_0 = 6$ if all pairs of the laser beams are contributed equally to the saturation process). According to Eq.(7) maxima of this function correspond to the minima of the ring radius. There are maxima of this function in terms of detuning and field intensity. The maximum in dependence on $\Delta$ means that there is a definite value of the detuning when ring radii acquire the minimal magnitude $r_{\min}$ (see Fig. 2).

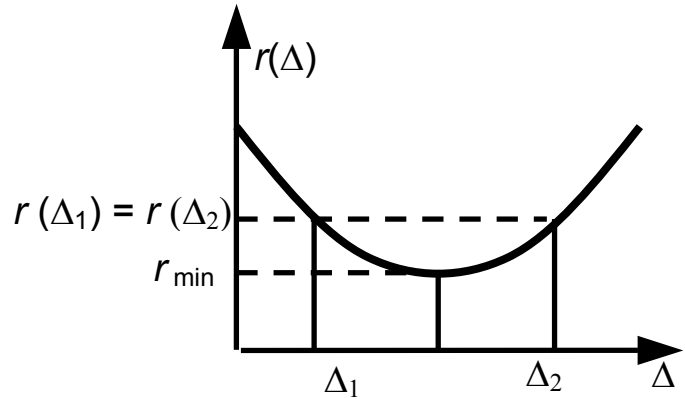

Fig.2. Dependence of the ring radius on the detuning $\Delta$. To the minimal radius value $r_{\min}$ corresponds the detuning $\Delta_0 = \sqrt{\frac{3}{20}(1 + s_0 I)}$.

The nonmonotonic dependence of the $\varphi$ − function on detuning means that due to the different dependences of the trapping (damping) forces and the vortex force on the detuning there is a certain frequency interval where to each value of the ring radius corresponds two values of the negative detuning ("frequency bistability"). The right branch of this curve-increasing of the ring radius on the detuning – was steadily established in Refs. [9-11]. For not so strong fields the threshold $\Delta_0^2 = \frac{3}{20}(1 + s_0 I)$ of the "frequency bistability" region is rather small ($\Delta_0 < 1$) and so for observation of the left branch one should ensure rather a great saturation, i.e. $I \gg 1$. However, then under no too large misalignment one may fall into region of ring instability (see Fig. 4 of Ref.[10]) where any ring structures are absent.

The general form of the function $\varphi(I,\Delta)$ is shown in Fig. 3 for two models of one and three pairs saturation (with $s_0 = 2$ and $s_0 = 6$).

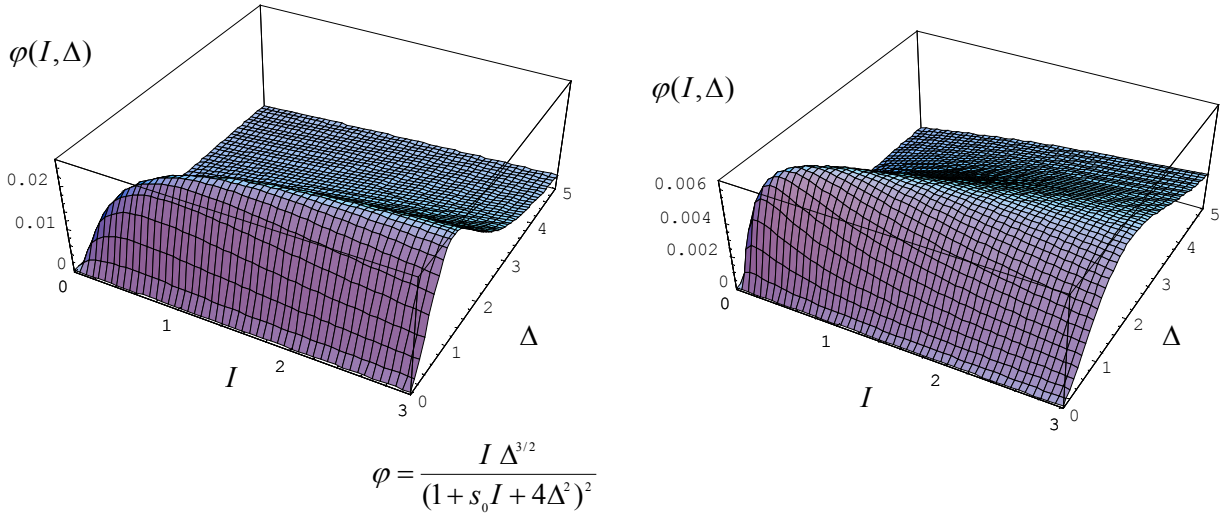

$$\varphi = \frac{I\,\Delta^{3/2}}{(1+s_0 I + 4\Delta^2)^2}$$

Fig.3. Surface of the function $\varphi(I,\Delta)$ for two saturation models with $s_0 = 2$ and $s_0 = 6$. The value of this function determines the magnitude of a ring radius in the interval from $r = r_{min}$ to $r = w$. Note that all points at the cross section of this function with the plane $\varphi(I,\Delta) = const$ correspond to the stable rings with the same radius. For the chosen model of saturation an absolute minimum of the value $r_{min}$ is determined by Eqs. (7), (10).

From Eq. (8) it is easy find an absolute maximum value of the function $\varphi(I,\Delta)$. It takes place under values

$$\Delta_0 = \sqrt{3}/2 \quad \text{and} \quad I_0 = 4/s_0 \qquad (9)$$

Then

$$\varphi_{max}(I_0,\Delta_0) = \frac{0.0403}{s_0} \ll 1 \qquad (10)$$

It means that to ensure rather noticeable dependence of ring radius one should maintain quite large value of the parameter $B$ in Eq. (7), which depends on the gradient of the magnetic field and the misalignment parameter $s$. In the conditions of the experiments [9-11] the value of this unitless parameter was of the order of $10^2$. It is sufficient for the noticeable variation of ring radius under detuning. .
Now we apply these results to consideration of ring structures in dual-isotope and two-species magneto-optical traps.

3. **About possibility of simultaneous rotation of the different isotopes in DIMOT along the same ring trajectory**

Simultaneous cooling and trapping of two rubidium isotopes was firstly realized by *Süptitz* et al [1] .Their schematic setup is shown in Fig.4.

**Simultaneous trapping of $Rb^{85}$ and $Rb^{87}$ by W.Suptitz et al (Opt.Lett.v.19,1571 (1994))**

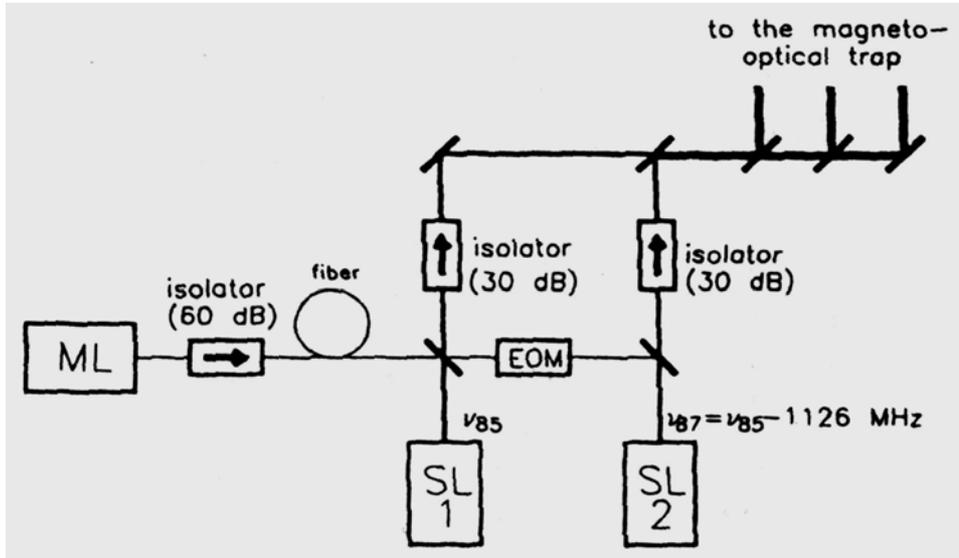

**ML- master laser SL(1,2) – slave diode lasers
EOM-electro-optical modulator.**

Fig.4. Schematic of the laser setup. The master laser (ML) light is injected in two slave diode lasers (SL 1,2) acting as cooling lasers, which are passively stabilized. The cooling transitions of the two rubidium isotopes $5S_{1/2} \to 5P_{3/2} (Rb^{85}: \ F=3 \to F=4; Rb^{87}: F=2 \to F=3)$ are different by 1126 MHz. The master laser is tuned to the cooling frequency of $Rb^{85}$. One slave laser is locked to the master laser without frequency shift, whereas the second slave laser is shifted by -1126 MHz by the electro-optical modulator. A beam splitter combines the light of two slave lasers to a two-frequency beam that cool both isotopes simultaneously. There are yet two free-running repumping lasers not shown at the setup (Fig. from Ref. [1]).

This setup was typically operated at the beams intensity $I \approx 5$, detuning $\Delta \approx 3$ and the magnetic-field vertical gradient $12 \ G/cm$. The achieved densities of the trapped isotopes in DIMOT were of the order of $10^{10} \ cm^{-3}$, i.e. comparable with densities in the conventional rubidium MOT. Since it was conventional MOT with the carefully aligned pairs of beams only flat-topped spatial core atomic structures were observed.

It is interesting to estimate the possibility of observation of the ring spatial structures in the *misaligned* DIMOT with the similar parameters of a trap. First of all let us estimate magnitude of the function $\varphi(I,\Delta)$ for the abovementioned values of the parameters $I$ and $\Delta$. For the models with $s_0 = 2$ and $s_0 = 6$ one obtains the magnitudes of the function $\varphi(I,\Delta)$ in the region of $12x10^{-3}$ and $4x10^{-3}$, respectively. With such values of the $\varphi$ − function one may hope to observe rings of the rotating atoms of rather small radii $r < w$ only under rather great values of the parameter: $B \sim 10^2 - 10^3$, i.e. under large misalignment and strong gradient of the magnetic field. Such regime of the co-rotating along the same trajectory atoms is shown in Fig. 5.

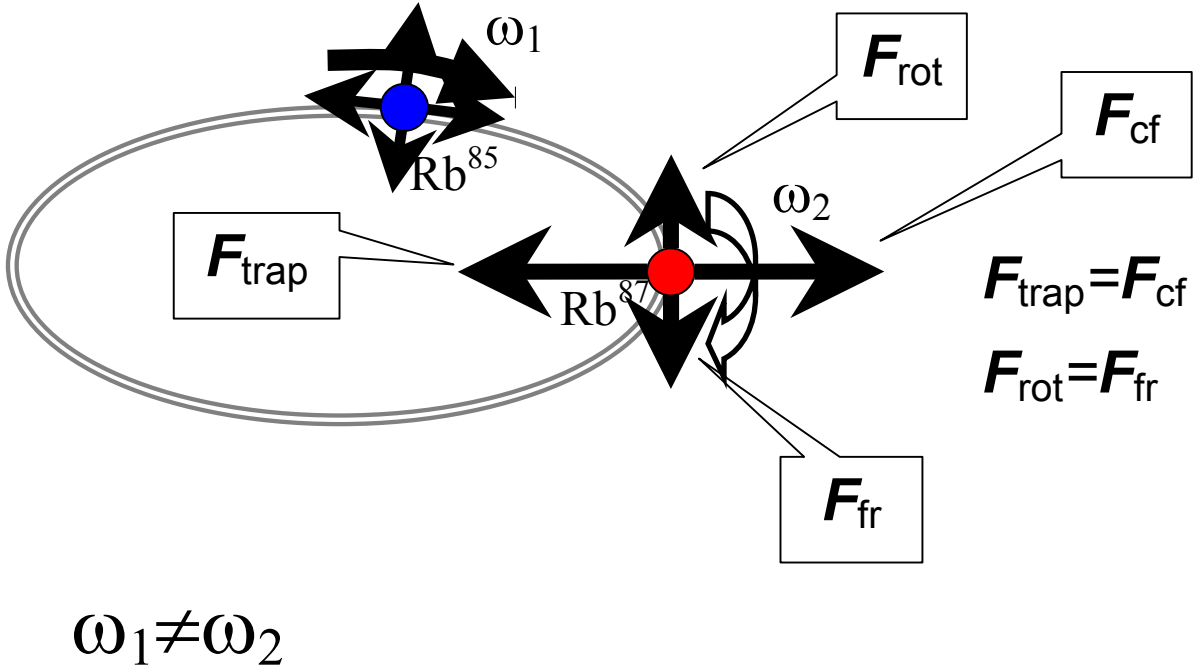

Fig.5. The imaginary setup of the co-rotating rubidium isotopes along the same ring of the radius $r$ trajectory. Since such rotations take place with the different angular velocities it leads to the motion of the isotope atoms with the relative velocity $\dfrac{|\omega_2 - \omega_1|}{r}$.

As it follows from setup in Fig. 4, presence of the electro-optical modulator allows to choose (regulate) the detuning and field intensities (parameters $\Delta$ and $I$) for each of the isotopes independently. Then the unique imposed criterion for equality of the radii of the atomic rings is condition

$$B_1 \varphi(I_1, \Delta_1) = B_2 \varphi(I_2, \Delta_2) \tag{11}$$

In the case of the rubidium isotopes the distinction in the ratio of the $B$-parameters $B_1/B_2 = \sqrt{m_2/m_1} \cong 1.01$ due to the difference of their masses can be neglected.
The physical sense of the condition (11) consists in the balance of the trapping and centrifugal forces in the radial direction and the vortex and damping (friction) forces in the azimuthal one.
In our oversimplified model of the isotropic harmonic trap neither of the forces depends on the angle of the rotation. In fact, due to anisotropy of the trap there is very strong azimuthal (on the angle $\varphi$) dependence of the vortex force $F_{rot}$:

$$F_{rot} \sim \left(e^{-r^2 \cos^2 \varphi} \cos^2 \varphi + e^{-r^2 \sin^2 \varphi} \sin^2 \varphi\right) \tag{12}$$

Fortunately, the friction force has exactly the same angular dependence. As a result, rotation of an atom occurs with constant angular velocity $\omega$. This fact turned out to be important under consideration of the beam imbalance and crosswind elevating force effects in the optical supermolasses setups [13,14].

## 4. Conclusion

In conclusion let us consider condition (11) in more details. The loci of this equation are shown in Fig. 6.

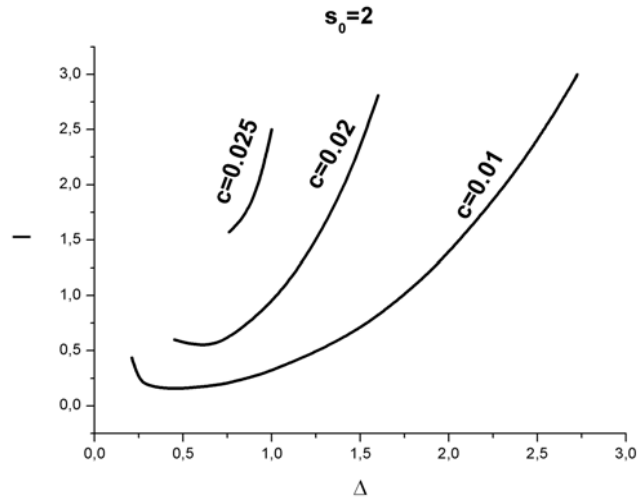

Fig. 6. Loci of equation $\varphi(I, \Delta) = c$ for several different values of the constant $c$. Any point (the pair of values $I$ and $\Delta$) on a given locus corresponds to the ring with a given radius.

Variant of the experiment with rubidium isotopes is considered here due to its relative simplicity because of possibility to employ for cooling only *one* master laser in six-beam geometry. Using *two* maser lasers one may cool any types of atoms and employ such collider setup for investigation of different atomic collisions under prescribed relative velocities. Note that one of the most vivid manifestation of the atomic collisions should be increase of thicknesses of two atomic rings with equal radii under their overlapping.